\documentclass[pre,twocolumn,aps,superscriptaddress,showpacs,floatfix]{revtex4}
\usepackage{graphicx}
\usepackage{dcolumn}
\usepackage{bm}

\begin{document}

\title{Interference of stochastic resonances; splitting of Kramers' rate}

\author{Pulak Kumar Ghosh$^a$,  Bidhan Chandra Bag$^b$
and Deb Shankar Ray$^a${\footnote {e-mail address:
pcdsr@mahendra.iacs.res.in}}}

\affiliation{$^a$Indian Association for the Cultivation of Science,
Jadavpur, Kolkata
700 032, India\\
$^b$Department of Chemistry, Visva-Bharati, Santiniketan 731 235,
India}

\noindent
\begin{abstract}
We consider the escape of particles located in the middle well of a
symmetric triple well potential driven sinusoidally by two forces
such that the potential wells roll as in stochastic resonance and
the height of the potential barrier oscillates symmetrically about a
mean as in resonant activation. It has been shown that depending on
their phase difference the application of these two synchronized
signals may lead to a splitting of time averaged Kramers' escape
rate and a preferential product distribution in a parallel chemical
reaction in the steady state.
\end{abstract}

 \maketitle
 The escape of a particle from a metastable state due to
thermal activation has been a major issue in chemical dynamics and
condensed matter physics for several
decades\cite{kram,hangi,ff2,san,db1}. As a typical paradigm in this
context, consider a
 Brownian particle in the middle well of a symmetric triple
well potential which diffuses symmetrically  to the left and the
right well. At a finite temperature and in absence of any bias force
the particles are activated only by inherent thermal fluctuation
resulting in equalization of population in the two side wells.
However, if, in addition, we allow the potential wells to roll by an
external periodic signal, the escape over the potential barrier is
modified by the interplay of the thermal fluctuations and coherent
external signal, due to \emph{Stochastic
Resonance}\cite{ben,luc,mcn,d41}. On the other hand when the height
of the potential barrier is made to oscillate symmetrically or
fluctuates around a mean value by the action of an external input
signal the mean escape time over the fluctuating barrier exhibits a
minimum at a particular value of frequency or correlation time of
the external source due to \emph{Resonant
Activation}\cite{dor,Bier,van,pk1}. In both of these cases the time
averaged escape rates from the middle well are equal and the
stationary population of the left and right wells remain the same.

Our aim of this letter is to explore a possible route leading to a
splitting of the time averaged Kramers escape rate from the middle
well due to the interference of these two resonances and to propose
a convenient method for controlling the pathways of a parallel
reaction for which the barrier heights corresponding to two product
states are equal. For example, take the case of nucleophilic attack
by $X^- $(a halide ion of HX) at the carboxyl group of a ketone,
say, $R_1(R_2)C=O$ producing $D-$ $R_1(R_2)C(OH)X$ and $L-$
$R_1(R_2)C(OH)X$, two optical isomers (enantiomers) having same
energy and stability but differing in their optical properties and
hence bio-chemical activities. The middle well of the potential
signifies the reactant state and the terminal wells represent the
two product states of the parallel reaction. Specifically, our
objective here is twofold: First, to understand how the asymmetry in
the time-averaged dynamics of the triple well potential driven
simultaneously by two sinusoidal forces results in differential
average escape rates to two product states and unequal distribution
of stationary population densities between them. Second, to explore
the role of phase difference of the two interfering forces in
determining asymmetric diffusion of the particles from the middle
well and resulting localization in one state. In other words we look
for a strategy for coherent control of pathways of a parallel
reaction. As an interesting offshoot of the analysis a selective
process of enrichment of one of the two isoenergetic isomers under
appropriate thermal condition is also explored.
\begin{figure}[bp]
\centering
\includegraphics*[width=8.2truecm]{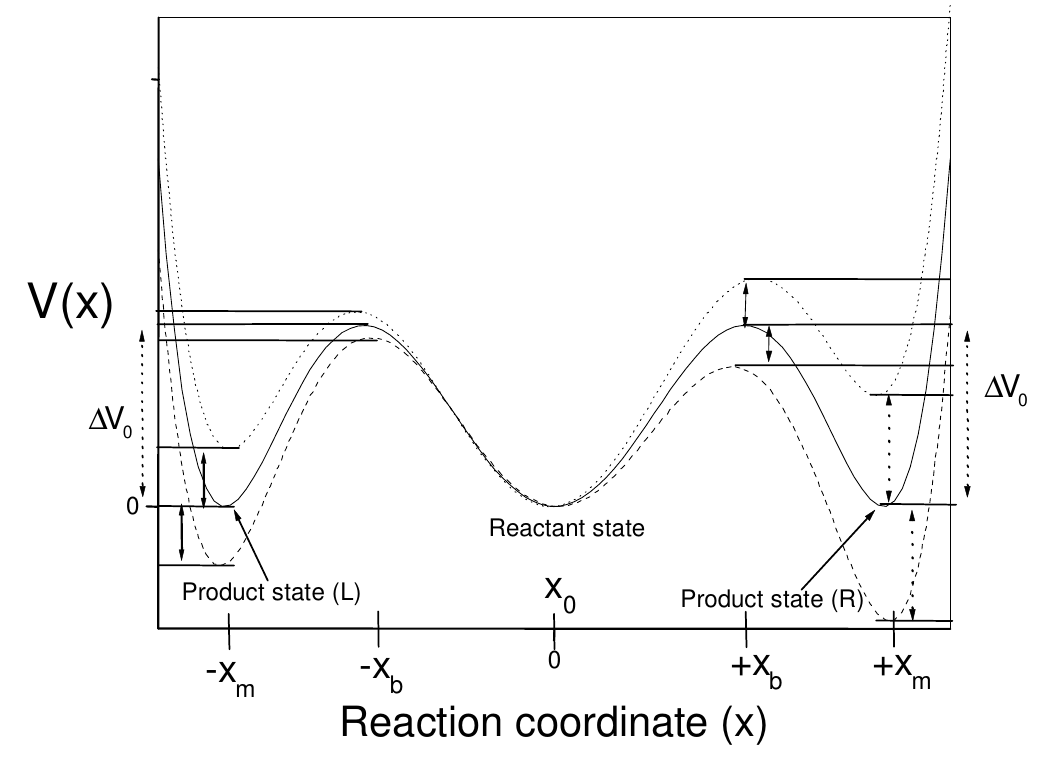}
\caption{(Color online) A schematic illustration the two
configurations of the potential under simultaneous action of the
two signals $a_1(t)$ and $a_2(t)$ for $\Delta\phi=\phi_1-\phi_2=0$
and synchronized frequencies $\omega_1=\omega_2$. }
\end{figure}

To illustrate the basic idea we begin with an overdamped Brownian
particle in a symmetric triple-well potential $V(x)$ (Fig.1) kept in
a thermal bath at temperature $T$ and subjected to two sinusoidal
signals $a_1(t)=A_1\sin{(\omega_1 t+\phi_1)}$ and
$a_2(t)=A_2\sin{(\omega_2 t+\phi_2)}$. The governing Langevin
equation is given by
\begin{eqnarray}\label{1}
\gamma\dot{x}= -V'(x)\;+\;A_1\sin{(\omega_1 t+\phi_1)}\;+ \;A_2
\;x\sin{(\omega_2 t+\phi_2)}+\Gamma (t)
\end{eqnarray}
where $V(x)=x^2(bx^2-c)^2$; $b$ and $c$ are the parameters of the
potential and $\gamma$ is the dissipation constant. $\omega_i$ and
$\phi_i$ ($i=1,2$) are the frequency and phase of the signals.
Thermal fluctuation of the bath is modeled by zero mean ($\langle
\Gamma(t)\rangle=0$) and delta correlation of noise, $\langle
\Gamma(t)\Gamma(t')\rangle=2 D \delta(t-t')$, $D$ being the strength
of the thermal fluctuation and is given by $D=kT/\gamma$. Here the
additive signal $a_1(t)$ rocks the potential wells sidewise, whereas
the multiplicative signal $a_2(t)$ sets a symmetric oscillation of
the barrier height around $\Delta V_0( = {4c^3}/{27b})$ with an
amplitude $\pm{A_2 x_b^2}/{2}$ at $\pm x_m$, respectively ( since
the fluctuation is space dependent the amplitude of fluctuation of
the barrier height around $\Delta V_0$ is $\pm{A_2(x_m^2-
x_b^2})/{2}$ for terminal wells to middle well), where $\pm x_b$ and
$\pm x_m$ are the coordinates of two barrier tops and two terminal
potential minima, respectively. The two configurations of the
potential under simultaneous action of the two signals $a_1(t)$ and
$a_2(t)$ are schematically illustrated in Fig.1 for
$\Delta\phi=\phi_1-\phi_2=0$ and synchronized frequencies
$\omega_1=\omega_2$. As shown the barrier height for the transition
from middle to right well fluctuates with an amplitude $\pm
\left({A_2x_b^2}/{2}+A_1x_b\right)$ whereas for the middle to left
well the amplitude of fluctuation of the barrier height
 is $\pm \left({A_2x_b^2}/{2}-A_1^0x_b\right)$. If the
external modulations
 $a_1(t)$ and $a_2(t)$ are small and very slow implying
$\Delta V_0 \gg A_2, A_1$ and the Kramers escape time ($1/r_k$) for
the unperturbed system is much smaller than the time period of the
external input signals ($1/r_k\ll 2\pi/\omega_1=2\pi/\omega_2$) one
may consider the expressions for the time-dependent transition rates
from the middle to left and right wells as follows
$W_M^L(t) =r_k \exp{[-\frac{\;A_1 x_b\sin{(\omega_1 t+\phi_1)}\;-
\frac{\;A_2\;x_b^2}{2}\sin{(\omega_2 t+\phi_2)}}{D} ]}\label{2}$;
$W_M^R(t) = r_k \exp{[+\frac{\;A_1 x_b\sin{(\omega_1 t+\phi_1)}\;+
\frac{\;A_2\;x_b^2}{2}\sin{(\omega_2 t+\phi_2)}}{D}]}$
, respectively.  Here $r_k=\frac{\omega_0\omega_b}{2\pi
\gamma}\exp{\left[-\frac{\Delta V_0}{D}\right]}$ is the transition
rate from the middle well for the unperturbed system;
$\omega_0,\;\omega_b$ are the frequencies corresponding to the
potential minimum ($x_0$) and barrier top ($x_b$), respectively.
Following McNamara and Wiesenfeld \cite{mcn} if we expand the
exponential term of the time dependent transition rate and keep the
leading terms up to second order, the time averaged transition rates
for $\omega_1=\omega_2$ and $\phi_1=\phi_2=0$ are given by
\begin{eqnarray}
\langle W_M^L(t)\rangle _t &=& r_k \left\{
1+\frac{1}{4D^2}\left(A_1x_b-\frac{A_2}{2}
x_b^2\right)^2\right\}\label{4}\\
 \langle W_M^R(t)\rangle _t &=& r_k \left\{
1+\frac{1}{4D^2}\left(A_1x_b+\frac{A_2}{2} x_b^2\right)^2\right\}
\label{5}
\end{eqnarray}
It is apparent from the above expressions that as a result of
interplay of two resonances the period averaged transition rates
from the middle to left and right wells significantly differ from
each other at very low temperature and tend to equalize in the high
temperature limit. In Fig.2 we present two representative plots for
the ratio of transition rate to right and left wells as a function
of temperature and compare the result with simulations in the
adiabatic regime $1/r_k\ll 2\pi/\omega_1=2\pi/\omega_2$.

\begin{figure}[bp]
\centering
\includegraphics*[width=8.2truecm]{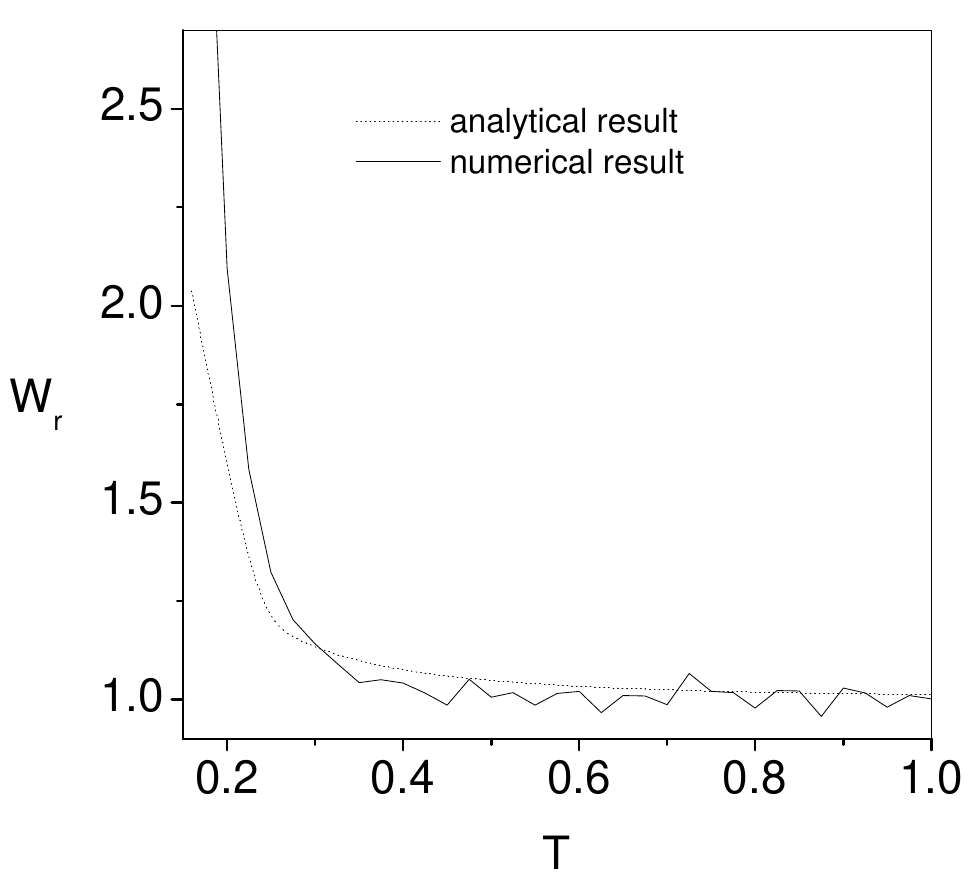}
\caption{(Color online) A comparison between analytical result(
based on analytical expression (2,3)) and numerical simulation
plotting $W_r$
 as a function of temperature for the
parameter set: $\Delta \phi=0,\; \omega_1=\omega_2=0.0005,
\;A_1=0.1, \;A_2=0.2,\; b=0.1,$ and   $c=1.0$}
\end{figure}

 We estimate
the ratio $W_r$($=\langle W_M^R\rangle/\langle W_M^L\rangle$) of the
transition rate as a function of external driving frequencies
(synchronized frequencies, $\omega_1=\omega_2$ ) by standard
numerical simulation of the Langevin equation (\ref{1}) using Huen's
algorithm. We allow $10,000$ test particles to move from middle well
in either direction and count the number of particles which arrive
in the left well($n_L$) and right well($n_R$), to calculate the
ratio of the transition as  given by $W_r=\langle
W_M^R\rangle/\langle W_M^L\rangle=n_R/n_L$. It is apparent from
Fig.2 that the numerical analysis matches fairly well with our
theoretical result. As a result of fluctuation of the barrier
heights for both left and right wells, the transition rates $\langle
W_M^L\rangle$ and $\langle W_M^R\rangle$ exhibit resonant activation
independently as expected (when the Kramers escape time coincides
with $2\pi/\omega_2$ ). Moreover as the amplitude of fluctuation of
the barrier height corresponding to the right well is larger (see
Fig.1), the ratio $W_r$ differs significantly from unity and
exhibits a resonance when the latter is plotted as a function of the
frequency under a phase matched condition $\phi_1=\phi_2$.

\begin{figure}[bp]
\centering
\includegraphics*[width=8.2truecm]{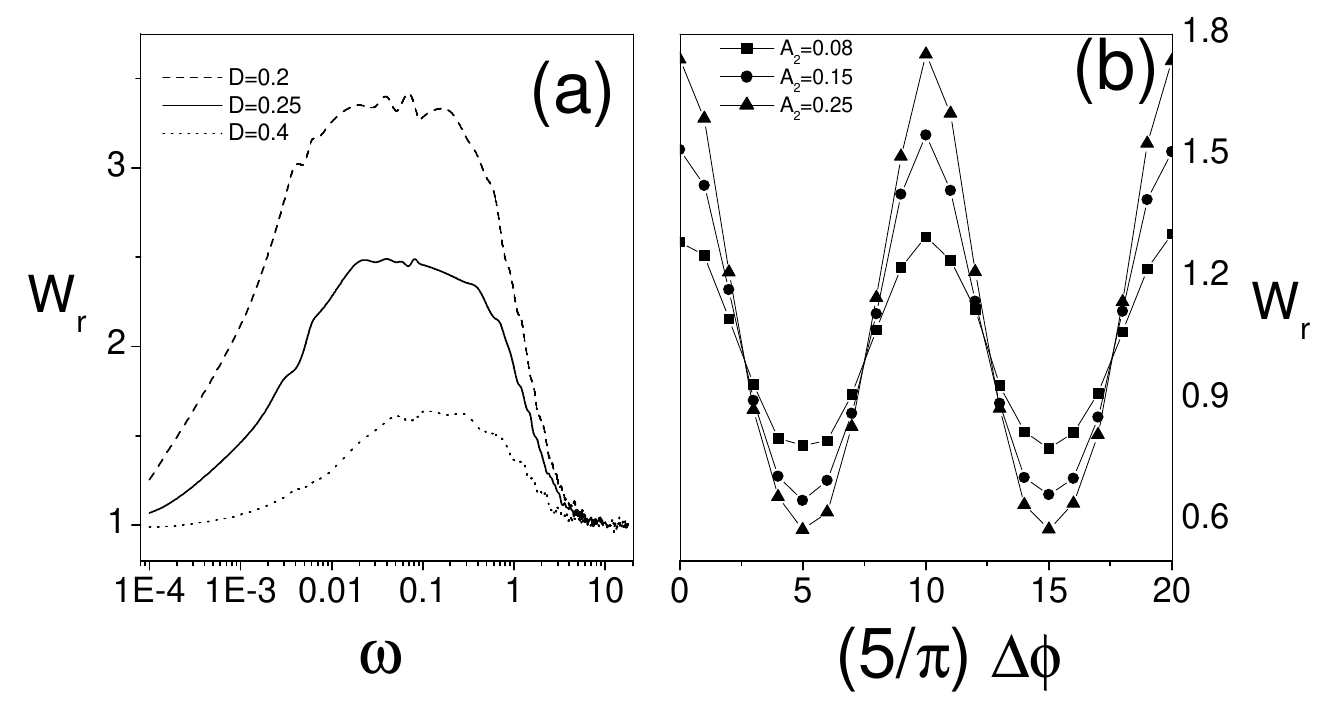}
\caption{(Color online)  (a) Variation of $ W_r$ as function of
frequency for different values of temperature $T=0.2$(dashed
line), $T=0.25$ (Solid line) and $T=0.4$ (dotted line) and for the
parameter set: $\Delta \phi=0, A_1=0.1, A_2=0.2, b=0.1, c=1.0 $.
(b) $ W_r$ vs phase difference $ \Delta \phi$ plot for different
values of $A_2$ and for the same parameter set as Fig.3(a) but for
$\omega_1=\omega_2=0.05$ and $T=0.2 $.}
\end{figure}

 In
Fig.3(a) we plot this ratio as a function of frequency between the
synchronized input signals for several values of temperature to
exhibit this asymmetry in resonant activation due to differential
transition rate. In Fig.3(b) we plot the ratio of the transition
rate as a function of the phase difference of the two input signals
for different values of strength of the input signal($A_2$). It is
observed that the transition rate towards the side wells are equal
($W_r=1$) for the phase difference $\phi_1-\phi_2=\pi/2$. The ratio
of the rates $W_r$ can be inverted by reversing the phase difference
from $\phi_1-\phi_2=0$ to $\phi_1-\phi_2=\pi$. Therefore by
controlling the phase of the input signals $a_1(t)$ and $a_2(t)$, it
is possible to manipulate the transition of particles from the
middle well to the product states and hence the course of the
parallel reaction.

The above analysis is based on kinetic considerations. It is also
worthwhile to turn our attention to the time averaged distribution
of the particles in the two wells and the associated aspects of
localization\cite{ff5}. This is in the spirit of stationary product
distribution of a parallel reaction. In this case we allow the
trajectories to evolve dynamically for long time starting from an
arbitrary initial position in the well. Numerical simulation results
show that the residential time distribution (Fig.4) bears a marked
asymmetry corresponding to a stochastic localization of the

\begin{figure}[bp]
\centering
\includegraphics*[width=8.2truecm]{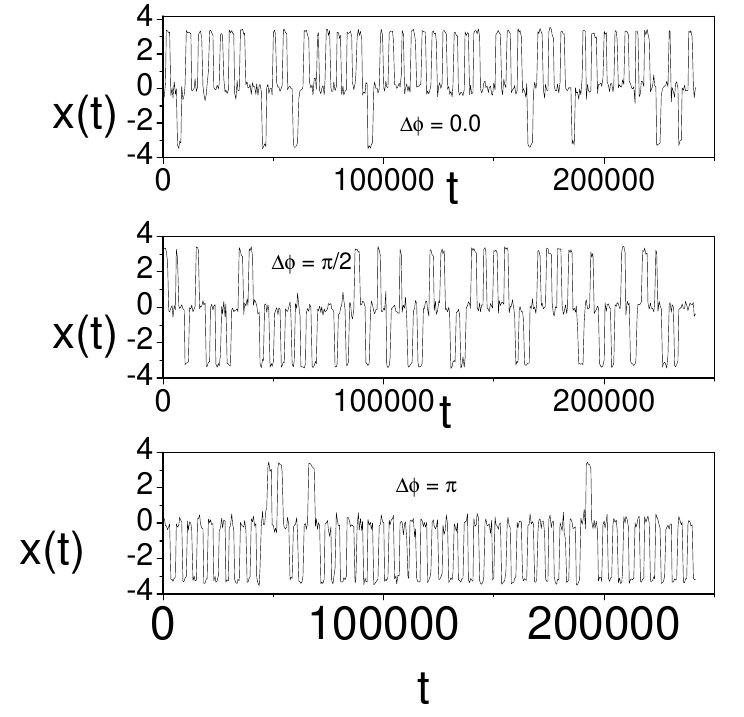}
\caption{(Color online)  Residential time distribution for several
values of phase difference ($\Delta \phi=0 $ (upper panel),
$\Delta \phi=\pi/2 $ (middle panel),$\Delta \phi=\pi $ (lower
panel)) between two input signals, for the parameter set:
$\omega_1=\omega_2=0.0013,\; A_1=0.1, A_2=0.25,\;T=0.155,\; b=0.1$
and $ c=1.0 $}
\end{figure}

particles in right well for $\phi_1-\phi_2=0$ and in the left well
for $\phi_1-\phi_2=\pi$. For $\phi_1-\phi_2=\pi/2$, the distribution
of $x(t)$ over time is more or less even for both the wells. In view
of the input signal synchronization, a qualitative interpretation of
this type localization  may be given as follows: So long as the
force $a_2(t)$ causing symmetric oscillation of the barrier height
attains its lower value, the tilting force $a_1(t)$ points to the
right well, so that the particle in the middle well move towards the
right well very quickly. On the other hand as the tilting force
points to the left $a_2(t)$ sets the barrier height at a larger
value and consequently the particle in the middle well takes
relatively larger time to speed up from middle to left well for the
simultaneous action of the synchronized signals. The particles in
the middle well therefore have a greater chance to cross the
right-hand barrier. In the right well, the amplitude of oscillation
of the barrier height is much larger and so the particle coming into
the right well escape from it more quickly and on returning back to
the middle well it has again two options to cross the barrier as it
was initially. Thus the particle dynamically spends most of the time
in between the middle and the right well. To proceed further we
require a quantifier which measures the asymmetry in localization in
the two wells. To this end we choose the mean position of the
particle as a measure of asymmetry. For a symmetric distribution
mean position $\langle x\rangle=0$ and for the localization of the
particles to left or right well, value of mean position is negative
or positive, respectively. In Fig.5(a) we present the variation of
mean position as a function of synchronized input signal

\begin{figure}[bp]
\centering
\includegraphics*[width=8.2truecm]{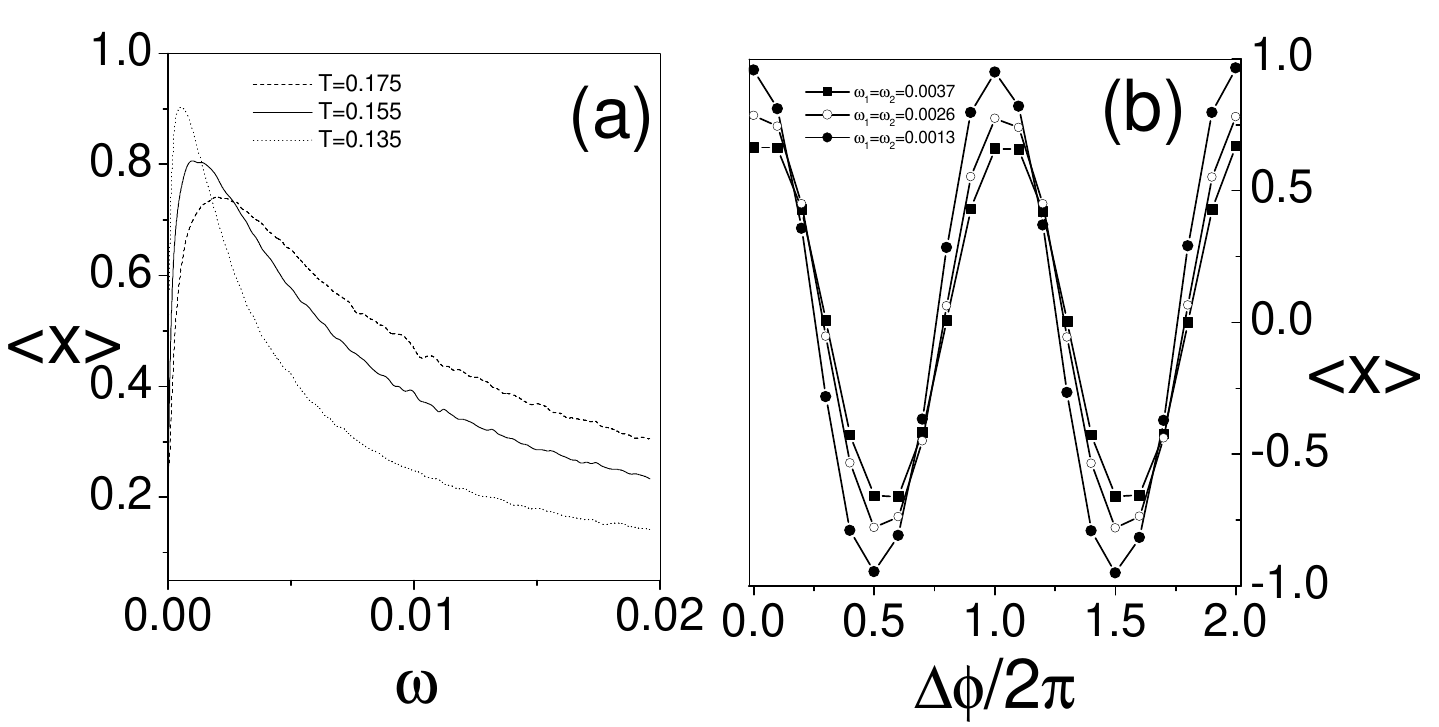}
\caption{(Color online) position($\langle x \rangle$) vs frequency
 plot for several values of temperature and for the
 parameter set: $\Delta \phi=0, \;A_1=0.1, \;A_2=0.25, \;b=0.1,
$ and $c=1.0 $. (b) Mean position($\langle x \rangle$) vs phase
difference ($\Delta \phi$) plot for different input signal
frequencies and for the same parameter set as Fig.5(a) but for
$T=0.155$.}
\end{figure}

frequencies. With increase of the input signal frequency the mean
position gradually shift to a maximum positive value followed by a
decrease to zero at high frequency. In Fig.5(b) we show the mean
position as a function of phase difference between the two input
signals. The mean position is zero for $\Delta \phi=\pi/2$ and it
departs from zero as the phase difference differs from $\pi/2$. For
a phase matched condition,  $\phi_1=\phi_2$ the particle are
localized in the right well while for a phase reversal $\Delta
\phi=\pi$, localization takes place in the left well.

\begin{figure}[bp]
\centering
\includegraphics*[width=8.2truecm]{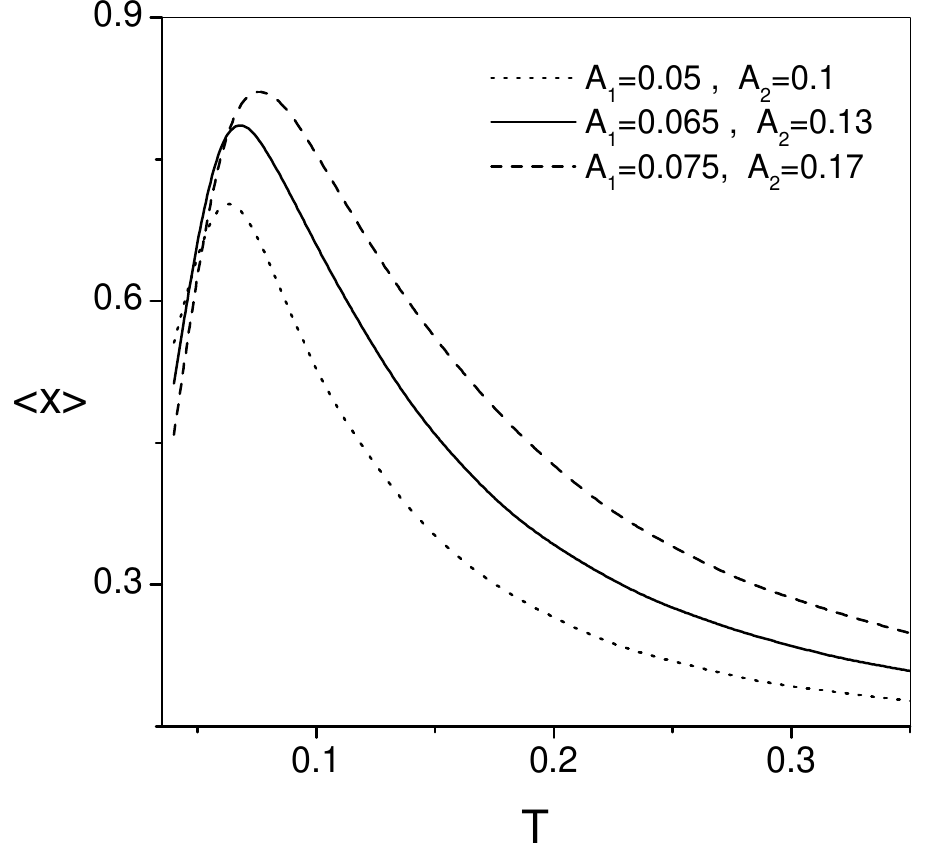}
\caption{(Color online)  Mean position <x> vs temperature plot for
different values of amplitude of the synchronized external signals
and for the parameter set $\Delta \phi$ =0,
$\omega_1=\omega_1$=0.0005, b=0.1, and c=0.6 (all the quantities
are dimensionless)}
\end{figure}

Can temperature influence the product distribution of a parallel
reaction at the steady state?  This question is intimately related
to the manipulation of incoherent condition rather than coherence in
selecting and controlling the reaction pathways. To have a closer
look into this aspect we examine the variation of mean position
$\langle x \rangle$ with temperature with the help of a discrete
three-state model for the triple well potential. Three states are
denoted by $x_0$, $\pm x_m$ for the symmetric unperturbed system
corresponding to three minima. The diffusional motion causes
transitions between them and it is schematically presented as
\begin{eqnarray}
\;\;\;\;k_{L}\;\;\;\;\;\;\;\;\;\;\;&\langle W_{M}^R\rangle& \nonumber\\
L\;\;\;\;  \rightleftharpoons\;\;\; M \;\;\;&\rightleftharpoons& \;\;\;\;R \nonumber\\
\;\;\;\;\langle W_{M}^L \rangle\;\;\;\;\;\;\;\;\;\;\;&k_{R}&
\nonumber
\end{eqnarray}
$k_L,\;k_R,\;\langle W_M^R \rangle ,\;\langle W_M^L \rangle$
  denote the time averaged rate of transition from left to
middle well, right to middle well, middle to right and middle to
left well, respectively. The  number of particles in  the three
states at time $t$
 are denoted by $n_{L},\; n_R$ and $n_L$. The governing master equations for
 $n_{i}$ ($i=L,\; R,\; M$) read as
 \begin{eqnarray}
\frac{dn_L}{dt}&=&-k_L\;n_L+\langle W_M^L \rangle\;n_M\\\label{6}
\frac{dn_R}{dt}&=&-k_R\;n_R+\langle W_M^R \rangle\;n_M\\\label{7}
\frac{dn_M}{dt}&=&k_L\;n_L+k_R\;n_R -(\langle W_M^L \rangle+\langle
W_M^R \rangle)\;n_M \label{8}
\end{eqnarray}
At the steady state ($\dot{n_L}=\dot{n_M}=\dot{n_R}=0$) the
probability of finding the particles at the three wells $P_{i}$
($i=L,\; R,\; M$) are $P_L=\langle W_M^L \rangle k_R/P,\;
P_R=\langle W_M^R \rangle k_L/P$ and $ \; P_M=k_L k_R/P$ where
$P=k_R \langle W_M^L \rangle+k_R k_L+k_L \langle W_M^R \rangle$. The
expression for the mean position is then given by
\begin{eqnarray}\label{9}
\langle x \rangle &=&\int_{-\infty}^{+\infty}x P(x)\;dx=x_m
P_R+x_0P_M-x_mP_L \nonumber
\\
&=&\left(\frac{\sqrt{27\Delta
V_0}}{2c}\right)\frac{\langle W_M^R \rangle k_L-\langle W_M^L
\rangle k_R}{k_R \langle W_M^L \rangle+k_R k_L+k_L \langle W_M^R
\rangle}
\end{eqnarray}
The above expression clearly shows the dependence of mean position
and probability on four time-averaged rate constants. Furthermore if
we assume that $k_R$ and $k_L$ do not differ significantly then
$\langle x \rangle$, in general, turns out to be positive since by
Eq(\ref{4}-\ref{5}) $\langle W_M^R \rangle$ is greater than $\langle
W_M^L \rangle$. Keeping in view of the Arrhenius temperature
dependence of the individual rate constants, the variation of

$\langle x \rangle$ with temperature is therefore expected to show a
bell-shaped curve. The departure of $\langle x \rangle$ from zero
towards positive direction indicates the preferential distribution
of the product in the right well. The numerical simulation of the
variation of mean position $\langle x \rangle$ as a function of
temperature for synchronized input signals under phase matched
condition as shown in the Fig.6 corroborates this assertion.

In summary, we have shown that depending on their phase difference,
an application of two synchronized signals on a particle in a triple
well potential may lead to a splitting of the time averaged Kramers
escape rate due to an interference of stochastic resonance and
resonant activation. This allows as to realize a strategy for
achieving a preferential product distribution in the steady state of
a parallel reaction. The present analysis thus reveals that in
stochastic energetics\cite{der} can be utilized to control
kinetically the pathways of a chemical reaction by appropriate
manipulation of coherence and/or inherent thermal condition.

\acknowledgments Thanks are due to CSIR, Govt. of India, for partial
financial support.


\begin{thebibliography}{99}
\bibitem{kram} H. A. Kramers, Physica {\bf 7}, 284 (1940).
\bibitem{hangi}P. H\"{a}nggi, P. Talkner, and M. Borkovec
Rev. Mod. Phys. {\bf 62}, 251 (1990).

\bibitem{ff2} S. Faetti, P. Grigolini and F. Marchesoni,
Z. Phys. B {\bf 47}, 353 (1982).

\bibitem{san} J. M. Sancho, M. San Miguel, S. L. Katz and J. D.
Gunton, Phys. Rev. A {\bf 26}, 1589 (1982).

\bibitem{db1} D. Barik, D. Banerjee and D. S. Ray, \textit{in Progress in Chemical Physics
Research}, Vol- 1, Edited by A. N. Linke, (Nova Publishers, New
York, ISBN: 1-59454-451-4, 2006).

\bibitem{ben} R. Benzi, G. Parisi, A. Sutera, and A. Vulpiani,
Tellus {\bf 34}, 10 (1982); R. Benzi, A. Sutera, and A. Vulpiani, J.
Phys. A {\bf 14} L 453 (1981).

\bibitem{luc} L. Gammaitoni, P. H\"{a}nggi, P. Jung and
F.Marchesoni, Rev. Mod. Phys., {\bf 70}, 223(1998).

\bibitem{mcn} B. McNamara and K. Wiesenfeld, Phys. Rev. A, {\bf
39}, 4854 (1988); B. McNamara, K. Wiesenfeld and R. Roy, Phys. Rev.
 Lett. 60, 2628 (1988).


\bibitem{d41} P. K. Ghosh, D.
Barik and D. S. Ray, Phys. Lett. A {\bf 342} 12 (2005).

\bibitem{dor} C. R. Doering and J. C. Gadoua, Phys. Rev. Lett. {\bf 69},
 2318 (1992).

\bibitem{Bier} M. Bier and R. D. Astumian, Phys. Rev. Lett. {\bf 71},
 1649 (1993); U. Z\"{u}rcher and C. R. Doering, Phys. Rev. E {\bf 47},
 3862 (1993).
\bibitem{van} C. Van den Broeck, Phys. Rev. E {\bf 47},
 4579 (1994).

\bibitem{pk1} P. K. Ghosh, D. Barik, B. C. Bag and  D. S. Ray, J. Chem. Phys.
{\bf 123}, 224104 (2005).
\bibitem{ff5} M. Borromeo and F. Marchesoni, Chaos {\bf 15}, 026110 (2005).
\bibitem{der} I. Der\'{e}nyi, M. Bier and R. D. Astumian, Phys.
Rev. Lett. {\bf 83}, 903 (1999); K. Sekimoto, J. Phys. Soc. Jap {\bf
66} 1234 (1997).
\end{thebibliography}
\end{document}